\documentclass[prb,twocolumn,amssymb,floatfix,showpacs,superscriptaddress]{revtex4}
\usepackage{epsfig}
\usepackage{amssymb}
\usepackage{natbib}
\usepackage{graphicx}
\begin{document}

\draft
\title{Structural and Magnetic properties of polymerized C$_{60}$ with Fe}
\author{A. Talyzin}
\author{A. Dzwilewski}
\affiliation{Department of Physics, Ume\aa~University, S-90187 Ume\aa,
Sweden}
\author{L. Dubrovinsky}
\affiliation{Bayerisches Geoinstitut, Universit\"at Bayreuth,
D-95440 Bayreuth, Germany}
\author{A. Setzer}
\author{P. Esquinazi}\email[E-mail address:]{esquin@physik.uni-leipzig.de}
\affiliation{Division of Superconductivity and Magnetism,
University of Leipzig, D-04103 Leipzig, Germany}

\date{\today}
\begin{abstract}
We provide evidence that high-pressure high-temperature (2.5 GPa and
1040~K) treatment of mixtures of iron with fullerene powders leads to the
complete transformation of iron into iron carbide Fe$_3$C. The comparison
of the magnetic properties (Curie temperature and magnetic moment) of the
here studied samples and those for the ferromagnetic polymer Rh-C$_{60}$
indicates that the main ferromagnetic signal reported in those samples is
due to Fe$_3$C and not related to the ferromagnetism of carbon as
originally interpreted. Taking into account the results obtained in this
study the original paper on ``Magnetic carbon" (Nature {\bf 413}, 716
(2001))  was recently retracted.
\end{abstract}

\maketitle

\section{Introduction}

The recently re-discovered ferromagnetism in carbon-based compounds is a
subject of actual interest in solid state physics.\cite{maka04} Magnetic
order at room temperature and above in a material with nominally only $s-$
and $p-$ electrons and without metallic ions is of importance for basic
research as well as for possible bio-compatible and spin-electronic
applications. Reports on the magnetic order observed in
pressure-\citep{makanat01,wood,naroz03} and
photo-polymerized\citep{maka03} fullerenes as well as in hydrofullerite
C$_{60}$H$_{24}$ \citep{antonov02} triggered a renaissance of the interest
in the magnetic response of carbon structures. The earlier experimental
study by \citet{murakami96}, who induced magnetic ordering in
C$_{60}$-crystals exposing them to light  from a xenon lamp in the
presence of oxygen, has been recently confirmed in
Ref.~\onlinecite{maka03}. Remarkable is the fact that the Curie
temperature measured by the authors in Ref.~\onlinecite{murakami96}
reaches 800~K.

The C$_{60}$ can be polymerized by square ring connections
(2+2cycloaddition mechanism) as a result of photo-irradiation, irradiation
with an electron beam or high-pressure high-temperature (HPHT) treatment .
Variety of one-, two- and three-dimensional structures have been reported
for polymerized fullerenes. According to the original reports,
polymerization at temperatures and pressures near the C$_{60}$  cage
collapse and graphitization of the 2D rhombohedral Rh-C$_{60}$ phase leads
to ferromagnetism with $T_C \sim 500~$K.\cite{makanat01} The  work of
\citet{wood} showed a maximum in the magnetization for samples prepared at
conditions near the cage collapse, though no clear information on the
impurities was provided in their publication. The studies done by
\citet{naroz03} on pressure polymerized fullerenes indicated a Curie
temperature higher than 800~K, clearly above that reported in
Ref.~\onlinecite{makanat01}. One would speculate that the density of
localized spins can vary from sample to sample and therefore, within a
mean field theory one might expect different Curie temperatures. However,
this difference added to the non negligible impurity concentration
\cite{hohne02,hancar03,spemann03} found in the samples from
Ref.~\onlinecite{makanat01} cast some doubts about the intrinsic nature of
the ferromagnetic signal. Therefore, the formation of iron-carbon
compounds cannot be ruled out.

The aim of this work is to show that HPHT treatment of fullerene powder
mixed with iron powder produces iron-carbide (Fe$_3$C, cementite). In this
study we compare the magnetic properties of the so prepared samples with
the previously reported ferromagnetism of Rh-polymer of C$_{60}$. We argue
that the formation of Fe$_3$C during the synthesis of the samples prepared
in Ref.~\onlinecite{makanat01} explains not only the 500~K Curie
temperature but also the absolute value of the magnetic moment of those
samples taking into account the measured Fe concentrations.

\section{Experimental}

Powders of C$_{60}$ (99.5\%, MTR Corporation) and iron (GoodFellow,
99.9995\% with nominal particle size $\sim 2~\mu$m) in different
proportions (three mass percents of Fe in sample 1 and ten mass percents
of Fe in sample 2) were gently mixed  in the agate mortar. Those powder
mixtures were loaded into a platinum capsule with tight lids pressed on
mechanically.  High-pressure high-temperature treatment was performed
using a standard  piston-cylinder system with a piston of $1/2$-inch
diameter.  An advantage of our piston-cylinder method is the relatively
large mass of samples (150-170~mg).  The loaded capsule was placed into a
standard high  pressure assemblage which consists of an alumina sample
holder inside a talc-pyrex assembly with  resistive heating provided by a
graphite tube.

 Synthesis of samples was performed at 2.5~GPa and 1040K  with heating
 time of 1000~s. According to literature data these  conditions favor
 formation of tetragonal polymeric phase of C$_{60}$.\cite{davy00}  It
 satisfies also the conditions of ferromagnetism reported in
 Ref.~\onlinecite{makanat01}:   the samples are synthesized just below the
 boundary of C$_{60}$ collapse.   Characterization of samples was
 performed by X-ray diffraction and Raman spectroscopy.    XRD patterns
 were recorded before and after HPHT treatment using Philips X'Pert
 and Bruker D8 powder diffractometers K$\alpha_1$-radiation in
 reflection mode. Silicon was used as an internal standard in some XRD
 runs. A Renishaw Raman 1000 spectrometer with a 514~nm excitation laser
 and a resolution of 2~cm$^{-1}$ was used in these experiments.

The magnetic properties of samples 1 and 2 were studied with a SQUID
 magnetometer from Quantum Design with RSO option. A pristine mixture of
 10\%Fe in C$_{60}$ powder (without HPHT treatment) was also studied as
 a reference (sample~3).  That powder was taken from the same batch as mixed
 for the sample~2.

\section{Results}
\subsection{Reaction of  iron powder with C$_{60}$\label{reaction}}

\begin{figure}
\includegraphics[width=8cm]{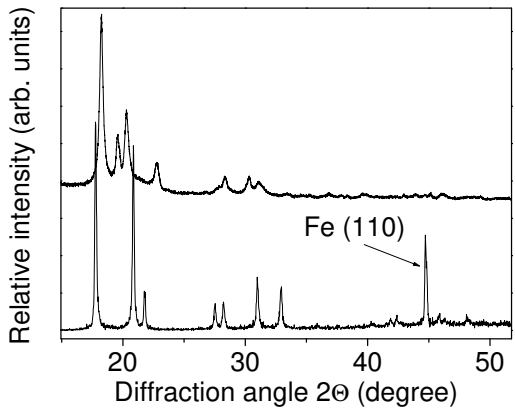}
\caption{XRD patterns recorded from the sample~2 (with 10\% of iron)
before (bottom) and after (top) HPHT treatment.} \label{XRD2}
\end{figure}

\begin{figure}
\includegraphics[width=8.0cm]{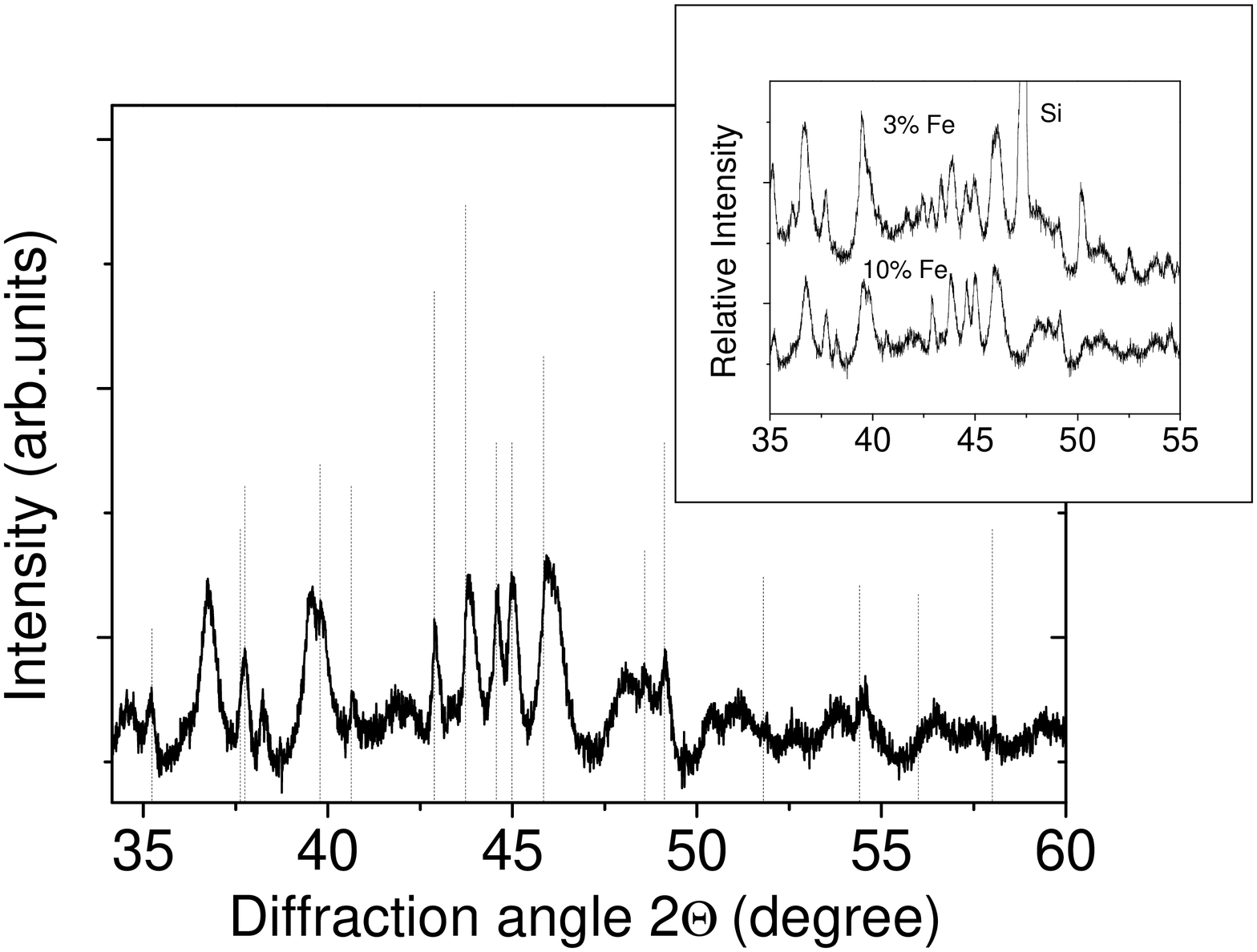}
\caption{Fragment of XRD recorded from the sample~2 after HPHT treatment.
Peak positions and relative intensity  from Fe$_3$C are marked with lines.
Other peaks are from polymeric phases of C60. The inset shows that XRD
patterns recorded from both samples~1 and 2  are in good agreement
(intensity is scaled for comparison).} \label{xrd2b}
\end{figure}

The XRD data of untreated C$_{60}$ and Fe powders show that no reaction
between these two components occurs after gentle grinding them in agate
mortar. Peaks from metallic iron were clearly observed in samples 1 and 2
(before treatment) together with peaks from usual fcc C$_{60}$ phase. No
peaks from iron oxides were detected confirming the high purity of
pristine powder. The XRD data for both samples after HPHT treatment showed
that C$_{60}$ powder polymerized into a tetragonal polymeric phase, the
same way as it is observed for pure C$_{60}$ at similar conditions.  At
the same time, the peaks from metallic iron  disappeared in HPHT treated
samples, see Fig.~\ref{XRD2}. XRD recorded   over longer period of time
revealed that iron reacted with C$_{60}$   forming iron carbide Fe$_3$C,
as seen in Fig.~\ref{xrd2b} for   both samples.\cite{xrd}

 Formation of iron carbide is possible only if part of C$_{60}$  molecules
collapses due  to reaction with iron. The mechanism of  reaction is not
known in detail, but some interesting  observations can be noted.  The
iron powder used in our experiments  was composed by grains with
relatively large size ($1-2~\mu$m).  Nevertheless, it is obvious from XRD
data that the iron carbide phase was  formed not just on the surface, but
whole grains were transformed into Fe$_3$C.  This process requires
diffusion of carbon from the surface to the core of  iron grains.
Formation of other carbon phases was not detected, which means  that most
of the carbon from the collapsed C$_{60}$ molecules  was consumed for the
formation of Fe$_3$C. Formation of Fe$_3$C is especially likely in the
temperature interval where C$_{60}$ molecules start to collapse or just
before this point. On one hand the  high temperature is required to
initiate  reaction of  Fe with C$_{60}$, on the other hand reaction with
collapsed fullerite (hard carbon phase) which forms above $\sim 1073~$K is
less likely for several reasons. Unlike C$_{60}$, graphite is a
thermodynamically stable modification of carbon with relatively low
chemical activity. Reaction of Fe with graphitic carbon starts at
significantly higher temperature (1273~K at 4-8~GPa),\cite{voca,tsu84}
while complete transformation of iron into Fe$_3$C (or Fe$_7$C$_3$ above
6~GPa) was reported at 1473~K-1523~K.\cite{tsu84,voca} Reactions of
carbide forming metals with C$_{60}$ were studied  previously only using
co-evaporation methods,\cite{hogb,taly} but it is likely that other metals
like Ti,V,Nb, etc., will also react with C$_{60}$ at the conditions of our
experiments. It should  be noted that Nb was one of the materials of the
sample container in previously  reported synthesis
experiments,\cite{maka04} which likely resulted in the collapse of the
C$_{60}$ at the surface layers of the sample reported in
Ref.\onlinecite{makasy03}.

\begin{figure}

\includegraphics[width=8cm]{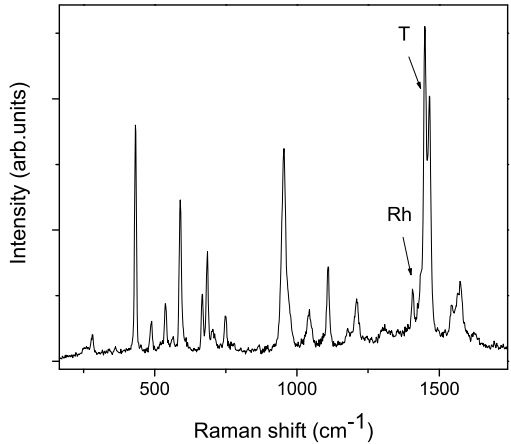}
\caption{Raman spectrum of  sample~2. Ag(2) mode peaks from Rhombohedral
(Rh) and tetragonal (T) polymers are marked.} \label{raman}
\end{figure}

   The results obtained by XRD were also confirmed by Raman spectroscopy
   (see Fig.~\ref{raman}).    The Raman spectra recorded in several points
   of samples~1 and 2 were in    good agreement with the spectra of
   tetragonal polymeric phase of C$_{60}$. The spectra recorded in several
   points of the studied samples     showed only slight changes in the
   relative intensity of some peaks.     Minor impurity of Rhombohedral
   polymer was also detected (see Fig.~\ref{raman},     peak at
   1407~cm$^{-1}$). The Fe$_3$C is difficult to detect by     Raman
   spectroscopy, but iron oxides (hematite and magnetite) would     be
   easy to identify taking into account the relatively large concentration
   of iron in sample~2.     Nevertheless, no traces of iron compounds were
   found in the     Raman spectra for both samples, which is in good
   agreement with XRD data. 

We conclude this section emphasizing our finding that metallic iron reacts
with C$_{60}$ at the HPHT conditions of our experiments. Finely dispersed
iron particles induce the collapse of some C$_{60}$ molecules with
formation of Fe$_3$C, while the majority of the fullerene molecules
transforms into the usual -- for these P-T conditions -- polymeric phases.
It is possible that the reaction described above can be found in mixtures
of other metals with fullerenes, therefore it can be proposed as a new
method for synthesis of bulk metal carbides in pure form. Formation of
Fe$_3$C was observed in our experiments at significantly lower
temperatures compared to reaction of Fe with graphite. It is also
relatively easy to remove fullerene from the C$_{60}$/Fe$_3$C composite
samples. The C$_{60}$ polymers can be depolymerized by annealing at 600K
and dissolved in toluene.

\subsection{Magnetic properties}

In what follows we present the raw data from the measurements of the
magnetic moments of the samples. In principle the total amount of
ferromagnetic phase in a sample is not necessarily equal or correlated to
its total mass ($\tilde{m}$). Therefore we have chosen to present data
using magnetic moment $(m)$ units as it was measured from the samples. In
the following discussion we will take into account the different masses of
the samples and the expected amount of ferromagnetic material.
Figure~\ref{300K} shows the hysteresis loops at 300~K for the three
samples prepared in this study. The inset shows the hysteresis of sample~1
in a restricted field range. Assuming that all iron in the samples 1 and 2
of Fig.~\ref{300K} would transform into Fe$_3$C after HPHT treatment, we
would have a mass {$\tilde{m}_{Fe_3C} \simeq 4.4 \times 10^{-4}$~g and
$\simeq 9.0 \times 10^{-4}~$g, respectively. Taking into account the
saturation magnetization of Fe$_3$C at room temperature $M_s \simeq
128~$emu/g,\cite{stab28} the expected magnetic moments at saturation due
to Fe$_3$C are $m_s \simeq 0.056~$emu and $\simeq 0.115~$emu for the two
samples, respectively. Both values are in agreement with the measured
curves within experimental error. For sample~3 of Fig.~\ref{300K} the mass
of Fe is $\tilde{m}_{Fe} \simeq 0.97~$mg. Taking into account that for
pure Fe at room temperature $M_s = 218~$emu/g, the expected magnetic
moment at saturation is $m_s \simeq 0.21~$emu, a value similar to the
measured one. These agreements also indicate the absence of giant magnetic
proximity effect\cite{coey02,cespedes04} between the ferromagnetic
particles and the carbon matrix, in agreement with the studies done in
Ref.~\onlinecite{hohprox}.

\begin{figure}
\includegraphics[width=8cm]{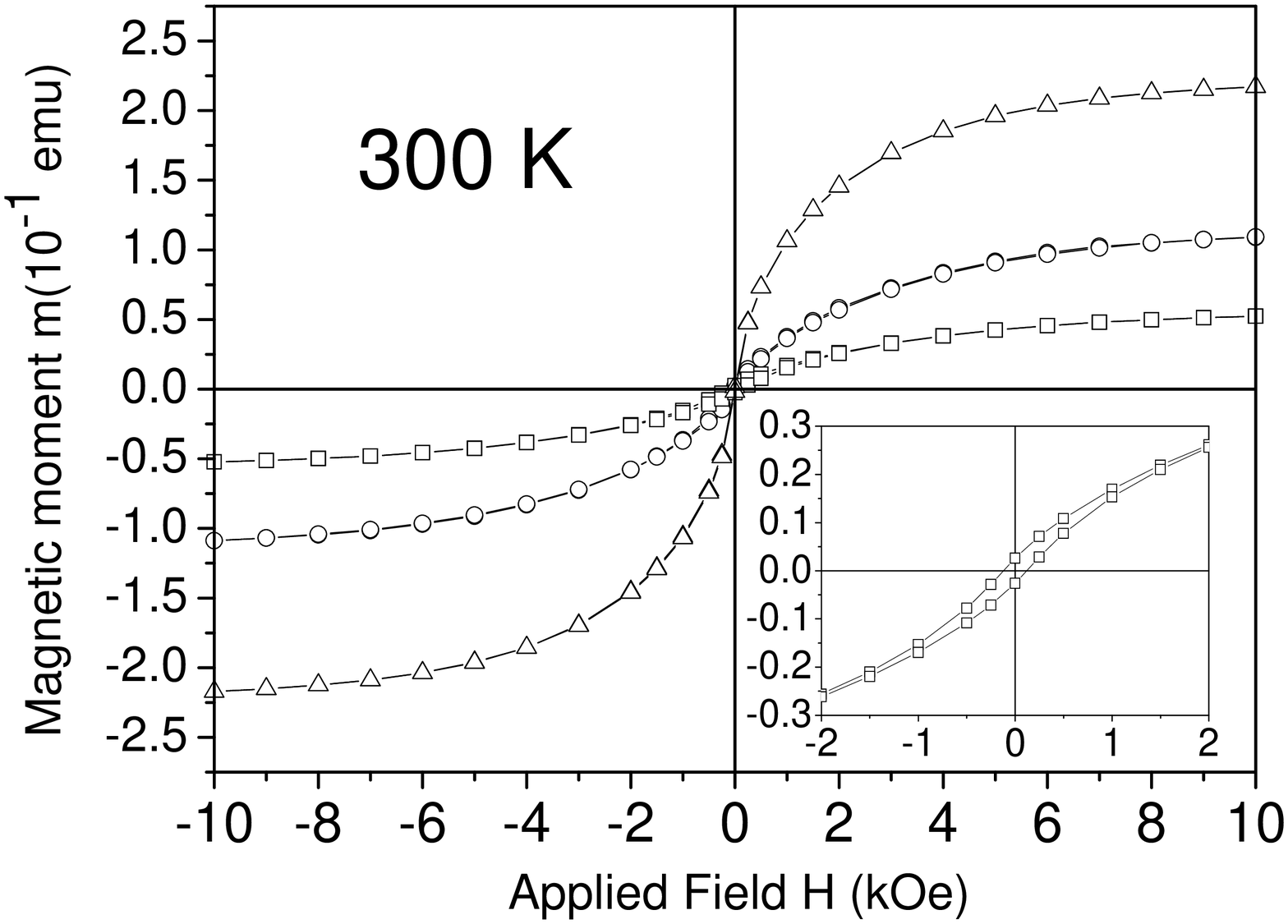}
\caption{Hysteresis loops of the magnetic moment measured for samples 1
($(\square)$, mass $\tilde{m} = 13.75~$mg), 2 ($(\circ)$, $\tilde{m} =
8.39~$mg) and 3 ($(\bigtriangleup)$, $\tilde{m} = 10.67~$mg) at room
temperature. The inset shows the hysteresis loop for sample~1 in a smaller
field region.} \label{300K}
\end{figure}

\subsection{Comparative analysis between C$_{60}$/Fe$_3$C composites
and ``magnetic carbon\cite{makanat01}"}

In this section we demonstrate that the magnetic properties of
C$_{60}$/Fe$_3$C samples are similar to those previously reported for
``magnetic carbon".\cite{makanat01,hohne02,hancar03,spemann03}  It was
shown in section~\ref{reaction}  that all iron introduced into the samples
prior (or in the process) of HPHT treatment transforms into Fe$_3$C. The
main question in the comparative analysis can be formulated as follows:
was the amount of Fe contamination in previously studied samples of
``magnetic carbon" sufficient to explain the observed ferromagnetic
signals?

-- It should be noted here that the amount of Fe impurity in the original
paper on ``magnetic carbon" was given only for pristine C$_{60}$
powder,\cite{makanat01} while the contamination introduced in process of
synthesis was not taken into account.  The relatively large amount of Fe
impurities in those samples was discovered after the publication of
Ref.~\onlinecite{makanat01} and  reported in
Refs.~\onlinecite{hohne02,hancar03,spemann03}. It can be also noted that
the only two other publications that apparently confirmed the existence of
ferromagnetism in HPHT C$_{60}$ polymers did not provide any impurity
analysis.\cite{wood,naroz03} Rigourously speaking they can not be
considered as a confirmation for intrinsic ferromagnetism in polymerized
fullerenes. It should be also clarified that only two samples of
``magnetic carbon" from one set of samples synthesized in 1998 were found
to exhibit a Curie temperature  $T_c \simeq 500~$K.  Below we discuss the
Fe contamination levels and the observed ferromagnetic signals for these
two samples in more detail.

-- Let us estimate the saturation magnetic moment  at room temperature
expected for the 3.2~mg polymerized fullerene sample studied in
Ref.~\onlinecite{makanat01}, taking into account the impurity
concentration. Particle Induced X-ray Emission (PIXE) measurements
indicated that the Fe concentration in similar samples was inhomogeneously
distributed within the penetration depth of this method ($\sim
30~\mu$m).\cite{hohne02,hancar03,spemann03} In average the magnetic
samples had an Fe concentration of the order of 400~$\mu$g/g with an
uncertainty of a factor of two or larger. If we assume that 400~$\mu$g/g
iron in carbon would transform into cementite, we expect
$\tilde{m}_{Fe_3C} \simeq 1.37~\mu$g and a magnetic moment at saturation
$m_s(300~$K,Fe$_3$C)$ \simeq 1.75 \times 10^{-4}~$emu. The measured
value\cite{makanat01} was $m_s \simeq 2.5 \times 10^{-4}$~emu. We note
that the iron concentration was not determined for that particular sample
(named E17 and produced at 6~GPa and 973K).\cite{makanat01} Similar
magnetic fullerene samples, however, showed the same behavior. PIXE
measurements of the sample E16 (2.5~GPa, 1123K) reported in
Refs.~\onlinecite{hancar03,spemann03}, before the polishing for MFM
measurements, showed the following Fe concentrations (in $\mu$g/g): (a)
wide-beam measurements: 541, 448 and 340 in three different regions, (b)
probed with a microbeam at one of the surfaces: 1.370, 100, 200, 16.000
and 100 (average over the specific surface 482), (c) at other surface:
630, 372, 52, 78 (average 502). The value 175~$\mu$g/g Fe written in
Refs.~\onlinecite{hancar03,spemann03} is the average measured in the
polished surface only. The saturation moment measured in this sample of
mass 2.2~mg at 300~K was $m_s \simeq 2.0 \times
10^{-4}~$emu.\cite{hancar03} A concentration of $710~\mu$g/g of Fe$_3$C
would provide this moment.

-- Let us discuss now the origin for the 500~K Curie temperature.
Figure~\ref{tran} shows the magnetic moment as a function of temperature
for the three samples and the raw data\cite{makanat01,hancar03} for the
E17 and E16 samples (right axis). This figure shows clearly that the
ferromagnetic transition at $\simeq 500~$K observed in those studies
agrees with that of cementite, which has a Curie temperature of 483~K (see
Ref.~\onlinecite{bozo53}, page 366). The differences in the shape and
width of the transitions between the samples is mainly related to the
difference in grain and stoichiometry distribution; larger applied fields
would enhance also the width of the transition. We conclude therefore that
the amount of metallic
 impurities  determined by PIXE in the original ``magnetic carbon" polymerized
 samples and the obtained Curie temperatures demonstrate
 that the magnetic signals observed
 in these samples should be assigned to contamination.

\begin{figure}
\includegraphics[width=8cm]{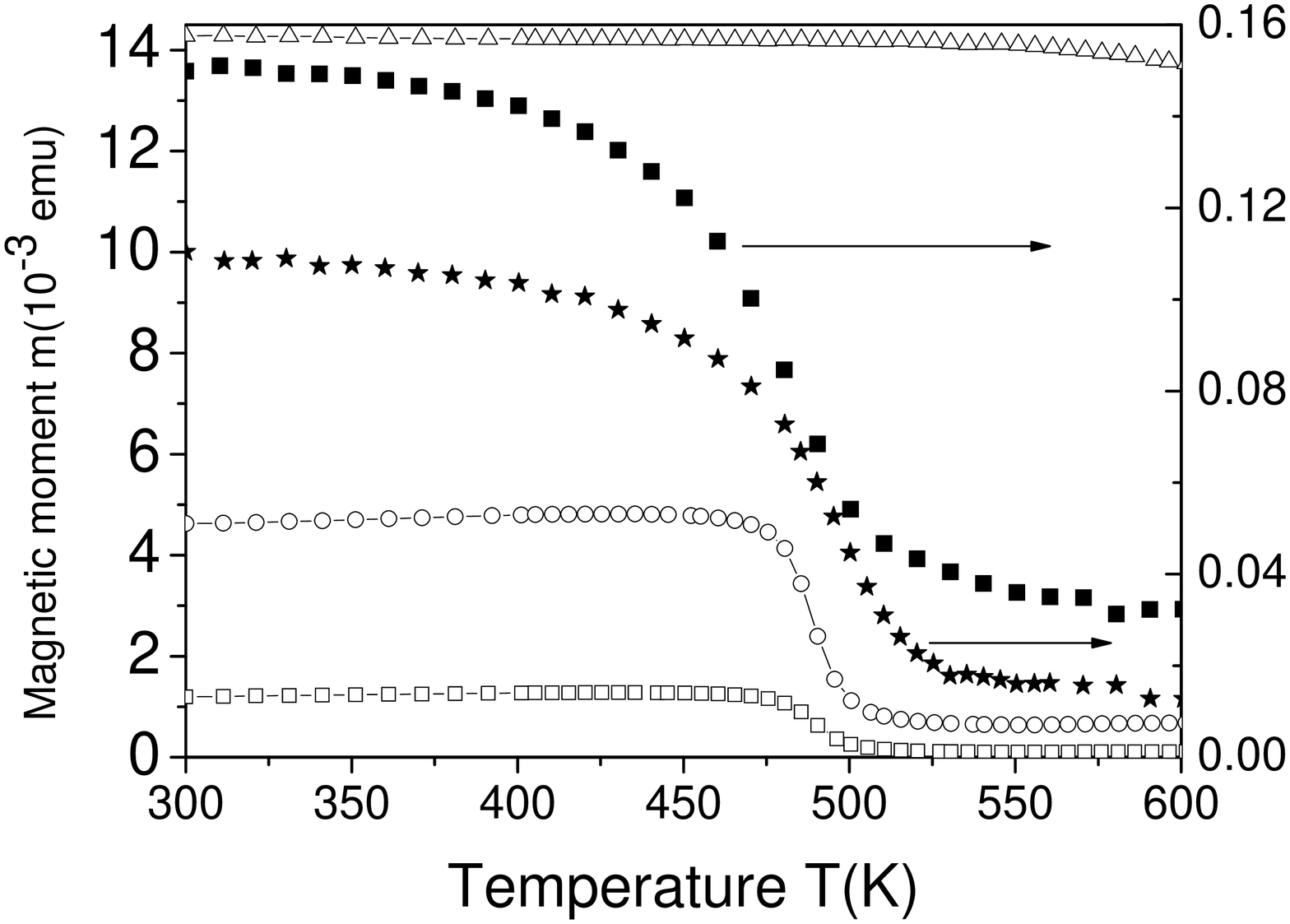}
\caption{Magnetic moment as a function of temperature for the three
samples studied in this work at an applied field of 500~Oe (the symbols
correspond to those in Fig.~\protect\ref{300K}). The sample mass was
$\tilde{m} = 2.0$, 2.0, 1.9~mg for samples 1, 2 and 3, respectively. Right
axis: raw data  $(\blacksquare)$ for the sample of mass $\tilde{m} =
3.2~$mg from Ref.~\protect\onlinecite{makanat01} and the sample E16 (mass
2.2~mg, $\bigstar$) from Ref.~\protect\onlinecite{hancar03} at the applied
field of 2~kOe.} \label{tran}
\end{figure}

\section{Discussion}

The report on intrinsic ferromagnetism of pure carbon in the form of
C$_{60}$ polymers produced by HPHT treatment\cite{makanat01} was based on
the following arguments:

1){\em  The contamination level of 22~ppm of the polymerized fullerene
samples was too small to account for the observed ferromagnetic moment.}
It is clear that only when contamination can be ruled out by a reliable
analysis (mainly of the elements Fe, Ni, and Co) the conclusion about an
intrinsic nature of the ferromagnetism in some new material can be done.
The level of impurities reported in the original paper\cite{makanat01} as
well as in a subsequent paper\cite{naroz03} was given only for pristine
C$_{60}$ powder. In some other reports about ferromagnetism in
C$_{60}$-based materials\cite{wood,makasy03,Boukhvalov} no analysis of the
magnetic impurities was presented at all. The contamination of the
material during the high pressure experiments was obviously not taken into
account. However, a later examination of one of the ferromagnetic samples
from Ref.~\onlinecite{makanat01} using PIXE showed a 10 to 20 times larger
concentration of ferromagnetic impurities than initially
reported.\cite{hohne02,hancar03,spemann03}  It can be noted that the
magnetic properties of the samples that exhibited ferromagnetism and a
Curie temperature of 500~K were discovered three years after their
synthesis in the year 1998. These samples were actually not intended for
studies of ferromagnetism and therefore the required precautions against
contamination were not considered at the time of their
synthesis.\cite{dav} Later studies on cleaner samples have not confirmed
the high levels of magnetization. The magnetization found in samples
specially synthesized with all possible precautions was on the level
0.004-0.0001~emu/g, which can be explained by less then 10~ppm of  Fe
contamination, if the Fe particles would have the ferromagnetism as bulk
Fe.\cite{han05,Boukhvalov}

2) {\em An unique Curie temperature of 500~K was assigned to ``magnetic
carbon".} The Curie temperature of 500~K was observed only for two
polymerized samples and can be naturally explained by the formation of
Fe$_3$C as it is shown in the presented study. The Curie temperature was
not measured on any other samples of ``magnetic carbon". The study
reported in Ref.~\onlinecite{naroz03} failed to find $T_c$ within the
temperature range of the used equipment (800~K). We may speculate that in
those samples some other contamination was involved, for example in the
form of magnetite or metallic iron (both with $T_c > 800~$K). The study
presented here confirmed that small particles of Fe ($\sim 2~\mu$m) are
completely transformed into Fe$_3$C at the conditions of  HPHT synthesis,
but such a transformation could be not complete if the Fe particles are
larger. The chemistry in real experiments can depend on many parameters,
as for example the particle size, time of the heat treatment and the
initial state of iron. For example, it was reported that an amount of
$80~\mu$g/g of Fe in C$_{60}$ was found to be in the form of hematite
already after the first preliminary part of the high pressure treatment
(pelletisation of powder).\cite{taly,mak04}  Starting from hematite, the
following chain of products could be obtained in the process of HPHT
treatment: hematite, magnetite, iron and iron carbide. The final reaction
product will strongly depend on the duration of the heat treatment and the
initial particle size. As a result, different Curie temperatures can be
observed in different HPHT treated samples. The amount of $80~\mu$g/g of
hematite reduced to pure iron would give a magnetization of $\simeq
0.018~$emu/g at saturation and in the case of complete transformation to
Fe$_3$C, 0.009 emu/g (similar to the ferromagnetic signals of a tetragonal
phase reported in Ref.~\onlinecite{makasy03}).

3) {\em Ferromagnetism was assigned only to special polymeric phases of
C$_{60}$.} In the first publication on the magnetic polymerized fullerene
sample, the rhombohedral phase has been attributed to be ferromagnetic.
 This assignment appeared to
be not true. All HPHT polymers of C$_{60}$ are known to depolymerized back
to pristine C$_{60}$ if heated above 550-600K.\cite{dwo97} The
ferromagnetism of ``magnetic carbon" was preserved after heating up to
640~K for two hours\cite{makanat01} and even after heat treatment at 800~K
for several hours\cite{hohne02} which means that the polymeric structure
was not anymore present in the samples after the first heating run and
therefore  responsible for the observed ferromagnetism.  We note that some
samples of the ``magnetic carbon" species of the first publication were
studied later and depolymerized as expected below 600~K.\cite{koro03}
Moreover, the same set of samples as in ref.~\onlinecite{makanat01} was
tested in an earlier study and was reported to depolymerize completely at
473~K.\cite{mak01} Finally, the recently published
corrigendum\cite{corrige05} confirmed that one of the two samples which
ever showed 500~K Curie temperature was synthesized at 2.5 GPa and 1125~K.
The temperature of 1125~K is well above the point of C$_{60}$ collapse.
Structural data for this sample were never published explicitly  despite
very detail characterization of magnetic properties by SQUID and
MFM.\cite{hohne02,hancar03,spemann03} This sample consisted largely of
graphite like carbon with  small fraction of tetragonal polymer and minor
impurity of rhombohedral.  Graphite like structure of this sample is
evident according to the conductivity
 measurements performed on the same sample and published prior to the original paper on
``magnetic carbon".\cite{mak01}

4) {\em Ferromagnetism was reported for samples synthesized only in a
``short temperature interval".} According to the original
publication\cite{makanat01} only samples prepared in the temperature
region of 1025~K to 1050~K were ferromagnetic. This assignment appeared to
be wrong as it follows from the published Corrigendum.\cite{corrige05}
Ferromagnetic loops shown in Ref.~\onlinecite{makanat01} appeared to be
obtained on the sample synthesized at 973~K, which was the lowest
temperature in the studied set (970~K-1170~K), while the second sample was
synthesized at 1125~K (above the point of C$_{60}$ collapse). It can also
be noted that a contamination with metallic impurities is most likely to
occur near the point of C$_{60}$ collapse.  A collapse of C$_{60}$ into a
more dense graphite-like hard carbon phase is also associated with a
significant volume decrease. Due to these reasons some pressure containers
were possibly cracked during experiments and contamination from outside
could penetrate into the reactive sample volume.

It can be  concluded that the existing evidence is not sufficient to
support a bulk intrinsic ferromagnetism in fullerene samples obtained by
high pressure high temperature treatment.  The claim of intrinsic
ferromagnetism in carbon samples can be made only when the contamination
with magnetic impurities is ruled out, but this is not the case for the
polymeric C$_{60}$ at the moment.  The study presented here clarifies the
last question that remained unanswered, namely how to explain a Curie
temperature of 500~K by contamination with magnetic impurities.

\section{Conclusion}
In summary, we have shown  that high-temperature high-pressure treatment
of fullerene samples intentionally mixed with iron powder before treatment
leads to the transformation of Fe into Fe$_3$C. The magnetic  data
obtained in this work compared with that from
Refs.~\onlinecite{makanat01,hancar03} show strong similarity, which
indicates that the {\em main} magnetic signal and the ferromagnetic
transition originally reported as intrinsic magnetism of carbon was likely
originated from Fe$_3$C. Taking into account the results obtained in this
study the original paper on ``magnetic carbon" \cite{makanat01} was
recently retracted.\cite{retraction} Although MFM data indicate the
existence of magnetic domains in pure regions of some samples produced
from C$_{60}$ by HPHT treatment\cite{hancar03,spemann03}, it is not
possible from those data to estimate their contribution to the total
magnetic signal. We note that recent experimental study\cite{han05}
performed on  samples prepared from fullerenes with lower impurity content
and after HPHT treatment showed vanishingly small bulk magnetization,
indicating that the pressure polymerization of fullerenes is not an
appropriate method to produce magnetic carbon.

\begin{acknowledgments}
We thank Catherine McCammon for the help with the high-pressure equipment.
Fruitful discussions and the support of Y. Kopelevich, R. H\"ohne, D.
Spemann and J.W. Taylor are gratefully acknowledge. High pressure
experiments were performed at the Bayerisches Geoinstitut under the EU
``Research infrastructures: Transnational access" Programme (505320
(RITA)-High Pressure). The financial support of the Deutsche
Forschungsgemeinschaft DFG (grant Es 86/11-1) and of the EU under FP6
project "Ferrocarbon" is acknowledged. A.D. thanks Helge Ax:son Johnsons
foundation for support.
\end{acknowledgments}


\end{document}